\documentclass[
aps,
prc,
superscriptaddress,
nofootinbib,
floatfix]
{revtex4}
\usepackage{graphicx,
longtable,
color,
array,booktabs,
bm}
\usepackage[T1]{fontenc} 

\begin{document}
\title{Analysis of $^{115}$In $\beta$ decay through the spectral moment method}
%
\author{			Joel~Kostensalo}
\affiliation{		Natural Resources Institute Finland, Yliopistokatu 6B, FI-80100 Joensuu, Finland}
\author{        	Eligio~Lisi}
\affiliation{   	Istituto Nazionale di Fisica Nucleare, Sezione di Bari, 
               		Via Orabona 4, 70126 Bari, Italy}
\author{        	Antonio~Marrone}
\affiliation{   	Dipartimento Interateneo di Fisica ``Michelangelo Merlin,'' 
               		Via Amendola 173, 70126 Bari, Italy}%
\affiliation{   	Istituto Nazionale di Fisica Nucleare, Sezione di Bari, 
               		Via Orabona 4, 70126 Bari, Italy}

\author{        	Jouni~Suhonen}	
\affiliation{		Department of Physics, University of Jyv\"askyl\"a, P.O. Box 35, FI-40014, Jyv\"askyl\"a, Finland}		
\affiliation{		International Centre for Advanced Training and Research in Physics (CIFRA), P.O. Box MG12, 077125 Bucharest-M\u{a}gurele, Romania	}
\medskip
\begin{abstract}
We analyze the $^{115}$In $\beta$-decay energy spectrum through the spectral moment method (SMM), previously introduced in the context of $^{113}$Cd $\beta$ decay. The  spectral moments $\mu_n$ are defined as averaged $n^{\rm th}$ powers of the $\beta$ particle energy, characterizing the spectrum normalization ($n=0$) and shape ($n\geq 1$) above a given threshold. For $^{115}$In, we consider three independent datasets characterized by different thresholds. We also consider three nuclear model calculations with two free parameters: the ratio of axial-vector to vector couplings, $r=g_{\rm A}/g_{\rm V}$,  and the small vector-like relativistic nuclear matrix element (NME), $s=s$-NME. By using the most recent of the three datasets, we show that the first few spectral moments can determine $(r,\, s)$ values in good agreement with those obtained by full-fledged experimental fits. We then work out the {\color{black} SMM} results for the other datasets. We find that, although $g_{\rm A}$ quenching is generally favored, the {\color{black} preferred} quenching factors may {\color{black} differ considerably depending on the chosen experimental} data and nuclear models. We discuss various issues affecting both the 
overall normalization and the low-energy behaviour of the measured and computed spectra, and their joint effects on the experimentally quoted half-life values. Further $^{115}$In $\beta$-decay data at the lowest possible energy threshold appear to be crucial to clarify these issues. 
\end{abstract}
\maketitle

\section{Introduction}
\label{Sec:Intro}

The quest for the rare process of neutrinoless double beta decay ($0\nu\beta\beta$) in candidate nuclei represents the most promising approach to reveal the fundamental nature of the $\nu$ field, either Majorana ($\nu=\overline\nu$)  or Dirac $(\nu\neq \overline\nu)$; see \cite{Agostini:2022zub} for a recent review and a vast bibliography. A worldwide 
experimental program is underway to push the sensitivity to $0\nu\beta\beta$ decay half-life values as high as $\sim 10^{28}$~y with ton-scale detectors
\cite{Agostini:2022zub,Adams:2022jwx}. 

This program is being paralleled by theoretical efforts to improve the calculations of 
$0\nu\beta\beta$ decay rates  in various nuclear physics models 
\cite{Agostini:2022zub,Cirigliano:2022oqy}. At present, such calculations are still subject 
to considerable theoretical uncertainties, affecting not only
$0\nu\beta\beta$ decay searches in different isotopes, but also 
other (observed) weak-interaction processes such as two-neutrino double beta ($2\nu\beta\beta$)
and single beta $(\beta)$ decay \cite{Ejiri:2019ezh}. Benchmarking models with a variety of
nuclear data and processes is crucial to improve the reliability of $0\nu\beta\beta$ rate calculations 
via a data-driven approach \cite{Cirigliano:2022rmf}.  

Among various sources of uncertainties, particular attention has been given
to the possible reduction (so-called quenching) of the 
effective axial-vector coupling $g_{\rm A}$ \cite{Suhonen:2019qcd,Barea:2013bz}
with respect to its vacuum  value $g_{\rm A}^\mathrm{vac}=1.276$ \cite{UCNA:2010les,Mund:2012fq}. 
Disparate quenching factors $q=g_{\rm A}/g_{\rm A}^\mathrm{vac}<1$ have been advocated to explain reduced
Gamow-Teller (GT) strengths with respect to model expectations,
in a number of $\beta$ and $\beta\beta$ decays and other weak processes{\color{black} --}see 
\cite{Suhonen:2017krv}  for a review. Various issues related to
the nature and size of quenching 
(as due to physical effects or missing model ingredients) are matter of
research and debate \cite{Cirigliano:2022rmf,Ejiri:2019ezh,Suhonen:2017krv,Ejiri:2019lfs,Gysbers:2019uyb}, and their clarification is crucial 
to improve  $0\nu\beta\beta$ calculations.

From a phenomenological viewpoint, it appears useful to investigate processes particularly sensitive to $g_{\rm A}$, such as highly forbidden non-unique $\beta$ decays. Indeed,
{\color{black} some electron spectra of forbidden $\beta$ decays turn out to} change very rapidly (both in normalization and in shape)
around values $g_{\rm A}\sim 1$, due to subtle cancellations among large nuclear matrix elements (NME) in various
nuclear models  \cite{Haaranen:2016rzs}. However, in the few cases where
data are available, it appears 
difficult to reproduce both the measured spectral shapes and 
their normalization with the same value of $g_{\rm A}$  \cite{Suhonen:2017krv}. 

Concerning spectral shapes alone, a well-studied example is represented by the fourth-forbidden $\beta$ decay of $^{113}$Cd, {\color{black} the} spectrum {\color{black} of which} was measured in detail in \cite{Belli:2007zza,Belli:2019bqp} and \cite{COBRA:2018blx}. 
The analysis performed in \cite{COBRA:2018blx} constrained $g_{\rm A}$ via the so-called 
spectrum-shape method (SSM) \cite{Haaranen:2016rzs,Haaranen:2017ovc,Kostensalo:2017jgw}  within three nuclear models: 
the microscopic interacting boson-fermion model (IBFM-2), the microscopic quasiparticle-phonon model (MQPM), and the interacting shell model (ISM). While a successful description of the $^{113}$Cd spectral shape was obtained for  quenched values of $g_{\rm A}$ \cite{COBRA:2018blx}, the normalization (in terms of the decay half life $t_{1/2}$) was not satisfactorily reproduced for the same $g_{\rm A}$. An independent approach to $^{113}$Cd decay has been recently carried out in \cite{DeGregorio:2024ivh} within the so-called realistic shell model (RSM), assuming no free parameter. 
In comparison with the $^{113}$Cd data of \cite{Belli:2007zza,Belli:2019bqp} ,
the authors of \cite{DeGregorio:2024ivh} obtain a {\color{black} reasonable} RSM description of the spectral shape with unquenched $g_{\rm A}$. 
However, the predicted
normalization (in terms of $10^{\log ft}$) remains a factor of $\sim 4$ away from the experimental value
(see Table~IV in \cite{DeGregorio:2024ivh}). 

The problem of reproducing {\color{black} simultaneously} the absolute normalization was tackled 
in the so-called
``enhanced'' (or revised) SSM \cite{Kumar:2020vxr,Kumar:2021euw}, by noting that the so-called  small vector-like relativistic nuclear matrix element ($s$-NME) played a crucial role in $t_{1/2}$ estimates, while being quite uncertain theoretically {\color{black} due to fundamental limitations of theoretical models which utilize limited model spaces}. 
Then, by taking the $s$-NME as a free parameter (together with $g_{\rm A})$ in data fits,  
a satisfactory description of $^{113}$Cd decay in terms of both spectrum shape and $t_{1/2}$ was achieved 
\cite{Kostensalo:2020gha}. 
The enhanced SSM was also recently applied to the 
fourth-forbidden $\beta$ decay spectrum of $^{115}$In, leading to a simultaneous 
measurement of its half-life and spectral shape, for model-dependent fitted values of $g_{\rm A}$ and $s$-NME 
\cite{Pagnanini:2024qmi}.   
It should be noted that 
the recent RSM calculations of $^{115}$In decay with no free parameter, analogously to $^{113}$Cd, 
can describe well the measured spectral shape but not its normalization, 
that remains a factor $\sim 7.7$ away from the data \cite{DeGregorio:2024ivh}. 
Altogether, these phenomenological results 
show that it is nontrivial to achieve a successful comparison of theoretical and experimental forbidden decay
spectra in absolute terms. Note also that shape and normalization issues are subtly connected in any estimate
of the decay half-life, that depends on {\color{black} exactly how the electron spectrum is extrapolated} below the observational energy threshold.

In this context, it may be useful to adopt methodological simplifications in comparing data and calculations, 
circumventing full-fledged data fits that are amenable only to
the expertise of experimental collaborations. Such fits often aim at estimating  $t_{1/2}$,
by mixing observed spectral features (above threshold) with theoretically 
extrapolated shapes (below threshold), while a more clear separation between data and models 
is desirable.  A step in this direction was taken in our previous work 
on $^{113}$Cd decay \cite{Kostensalo:2023xzu} by introducing the 
so-called spectral moment method (SMM), briefly reviewed in the next section. 
The SMM exploits the property that a continuous spectrum, defined in a finite interval of its argument, can be discretized in terms of its moments $\mu_n$, namely, of the average values of the $n$-th power of the argument
\cite{Feller91,Shohat70,Schmudgen17}.
The zeroth moment $\mu_0$ defines the spectrum normalization, while $\mu_1$ and higher
moments parametrize its shape. Actually, for sufficiently smooth spectra, 
the basic information is contained just in the first few moments \cite{Akhiezer20,Talenti87}. 

As discussed in \cite{Kostensalo:2023xzu}, by applying the SMM to
the $^{113}$Cd data of \cite{Belli:2007zza,Belli:2019bqp}, interesting results are readily obtained 
by equating the experimental and theoretical values of a few spectral moments. In
particular, in the planes charted by the free parameters $g_{\rm A}$ and $s$-NME for the various nuclear
models, isolines of $\mu_0$ correspond to ellipses, while  isolines 
of $\mu_1$ and
higher moments correspond to hyperbolas. The intersections of these conic curves 
provided a simple understanding \cite{Kostensalo:2023xzu}
of more detailed results obtained through refined data fits \cite{Kostensalo:2020gha}. 
Moreover, since all moments are consistently defined above the experimental
threshold within the SMM, no assumption about low-energy spectrum extrapolations is needed (as instead
required by $t_{1/2}$ priors). While the latter issue is not particularly relevant for the
$^{113}$Cd spectrum of \cite{Belli:2007zza,Belli:2019bqp}  (that can be extrapolated 
below threshold almost ``by eye''), it may be of some importance in other nuclei such as $^{115}$In, 
where the low-energy shape is rather uncertain, as discussed below.

In this work{\color{black},} we systematically apply the SMM to the $^{115}$In decay spectrum, in order to compare 
the most recent measurement \cite{Pagnanini:2024qmi} with earlier ones at different thresholds 
\cite{Leder:2022beq,Pfeiffer:1979zz}, 
within the IBFM-2, MQPM, and ISM nuclear models \cite{Haaranen:2017ovc,Kostensalo:2020gha}. As in 
\cite{Kostensalo:2023xzu}, we assume two free parameters
($g_{\rm A}$ and the $s$-NME), that are determined by the intersection of 
two moment isolines ($\mu_0$ and $\mu_1$) and, to some extent, 
by the spread of higher moment intersections. We highlight the fact that different datasets entail 
noticeable normalization differences above threshold. 
The datasets are also compatible with different (and partly conflicting) low-energy spectral shapes, qualitatively covering all cases: from an increase, to a plateau, or even to a rapid decrease.  
These spectral differences, partly compensated in $^{115}$In half-life estimates for accidental reasons,
emerge rather clearly within the SMM, e.g., in terms of  
$g_{\rm A}$ and $s$-NME parameter values in the various cases considered. 
In particular, while $g_{\rm A}$ quenching is generally favored in all cases, its
quantification suffers from spectral ambiguities at low energy, that remain unresolved by using 
the available data. 
Further and accurate $^{115}$In decay spectrum measurements
with the lowest possible experimental threshold would be highly desirable to clarify some residual
spectral issues. 
To our knowledge, such findings have not been discussed before in the literature.

Our work is structured as follows. In Sec.~\ref{Sec:SMM} we briefly 
describe the SMM introduced in \cite{Kostensalo:2023xzu}, in order to set the notation.
In Sec.~\ref{Sec:Data} we discuss and compare the input $^{115}$In data, as taken or elaborated from
the observations reported in refs.~\cite{Pagnanini:2024qmi,Leder:2022beq,Pfeiffer:1979zz}.
In Sec~\ref{Sec:Models} we discuss the theoretical formalism and the
parameter space of nuclear models in terms of $g_{\rm A}$ and $s$-NME.
In Sec.~\ref{Sec:Comparison} we compare in detail data and calculations in terms
of the SMM. We achieve a very good understanding of the parametric
results reported in \cite{Pagnanini:2024qmi} and a reasonable understanding of those reported 
in \cite{Leder:2022beq}, while we can only roughly capture the oldest results in \cite{Pfeiffer:1979zz}.
We also compare and discuss the normalization and shape variants 
emerging from the SMM analysis of different data and models, and highlight current uncertainties 
in the low-energy behaviour of the decay spectrum.
In Sec.~\ref{Sec:Summary} we summarize the results and discuss the perspectives of our work,
also in the light of other forbidden decay spectra that might be studied via their moments.

\section{The spectral moment method (SMM)}
\label{Sec:SMM}

The $\beta$-decay energy spectrum $S(T_e)$ corresponds to 
the fractional number of decays $n_e$ per single nucleus and per unit of time $t$ and of 
$\beta$ kinetic energy $T_e$, observed from a given threshold 
$T_\mathrm{thr}$ up to the maximum $Q$-value:
\begin{equation}
\label{spectrum}
S(T_e) = \frac{d^2n_e}{ dt\, dT_e}\ , \ T_e \in [T_\mathrm{thr},\,Q]\ .
\end{equation}
We shall use as units $[T_e]=\mathrm{keV}$ and $[t]=\mathrm{s}$ (or $\mathrm{y}=3.156\times 
10^{7}~\mathrm{s}$). 
For $^{115}$In, the endpoint is accurately measured as $Q=497.489$~keV \cite{Blachot:2012tby}. Note that in \cite{Kostensalo:2023xzu} we 
expressed $S$ in terms of the dimensionless total energy $w_e=1+T_e/m_e$, as
used in theoretical calculations \cite{Haaranen:2016rzs,Haaranen:2017ovc}. In this work we prefer to keep the dimensional 
argument $T_e$ for both
theoretical spectra (denoted as $S^t$) and experimental spectra ($S^e$), in order   
to facilitate contact with various published data on
$^{115}$In decay. Our results do not depend on such conventional choice.

The spectral moment method (SMM) discretizes the information contained in the 
continuous function $S(T_e)$ in terms of its moments $\{\mu_n\}_{n\geq 0}$.  The zeroth moment,
\begin{equation}
\label{moment0}
\mu_0 = \int_{T_{\rm thr}}^{Q}S(w_e)\,dT_e\ ,
\end{equation}
represents the decay rate per nucleus above the energy threshold, fixing the spectrum normalization.
The $n\geq 1$ moments, defined as
\begin{equation}
\label{moment}
\mu_n = \frac{\int_{T_{\rm thr}}^{Q}S(T_e)\,T_e^n\,\,dT_e}{\int_{T_{\rm thr}}^{Q}S(T_e)\,dT_e}\ (n\geq1)\ ,
\end{equation}
embed instead the (normalization independent) spectrum shape information above threshold. The corresponding units are $[\mu_0]=\mathrm{s}^{-1}$ and
$[\mu_n]=\mathrm{keV}^n$ for $n\geq 1$. Note that in \cite{Kostensalo:2023xzu}
the moments were defined in terms of the parameter $w_e$ and were thus dimensionless for $n\geq 1$.

We remark that $\mu_0$ corresponds to the total decay rate only for
null threshold ($T_\mathrm{thr}=0$), in which case it provides the
decay half life $t_{1/2}$ via
\begin{equation}
\label{rate}
T_\mathrm{thr}=0\ \rightarrow\ \mu_0 = \frac{\ln 2}{t_{1/2}} = \int_0^{Q} S(T_e)\,dT_e\ .
\end{equation}

In general, any smooth spectrum $S(T_e)$ can be described by the series of its moments, 
$\{\mu_n\}_{n\geq 0}$ \cite{Feller91,Shohat70,Schmudgen17}. In practice, for sufficiently smooth shapes,  the most 
relevant features of $S(T_e)$ can be captured by a truncated series with relatively small $N$,
$\{\mu_n\}_{0\leq n\leq N}$  \cite{Akhiezer20,Talenti87}. 
The comparison of experimental spectra $S^e$
with some model predictions $S^t({p})$ (where ${p}$ represents the free parameters of the model)
can then be approximately reduced to the comparison of a few experimental moments $\{\mu^e_n\}_{0\leq n\leq N}$ with the corresponding theoretical values $\mu_n^t({p})$. The essence of the SMM is the reduction of continuous spectral information to a few discrete quantities (the moments).

The  SMM was applied in \cite{Kostensalo:2023xzu} to study the $^{113}$Cd $\beta$-decay spectrum and its implications, e.g., for the quenching of $g_{\rm A}$. It was shown that interesting results could be obtained already
by a quantitative analysis of $\mu_0$ and $\mu_1$, supplemented by qualitative considerations about higher moments (up to, say, $N=6$), without the need for refined data fits. In the following, the SSM approach will be applied to the $^{115}$In $\beta$ decay spectrum, as measured by three different experiments.

\vspace*{-3mm}
\section{$^{115}\mathrm{\bf In}$ spectrum: Input experimental data}
\label{Sec:Data}
\vspace*{-1mm}

In this section we describe our input $^{115}$In spectrum data 
as taken or elaborated from three
different datasets, dubbed as: AC24 (ACCESS 2024 results) \cite{Pagnanini:2024qmi}, LE22 
(Leder {\em et al.}  2022 results) \cite{Leder:2022beq}, and PF79 (Pfeiffer {\em et al.} 1979 results) \cite{Pfeiffer:1979zz}. 
In these experiments, by using different techniques, the decay signals were observed above different thresholds
and {\color{black} then} processed in the data analysis. 
The three datasets discussed below should not be considered as official experimental data, but as
our best description of the signal spectra in terms of normalization and shape, 
that we have been able to recover on the basis of published results, plus supplementary
information when available from the collaborations. 

\begin{figure}[t]
\begin{minipage}[c]{0.7\textwidth}
\includegraphics[width=0.72\textwidth]{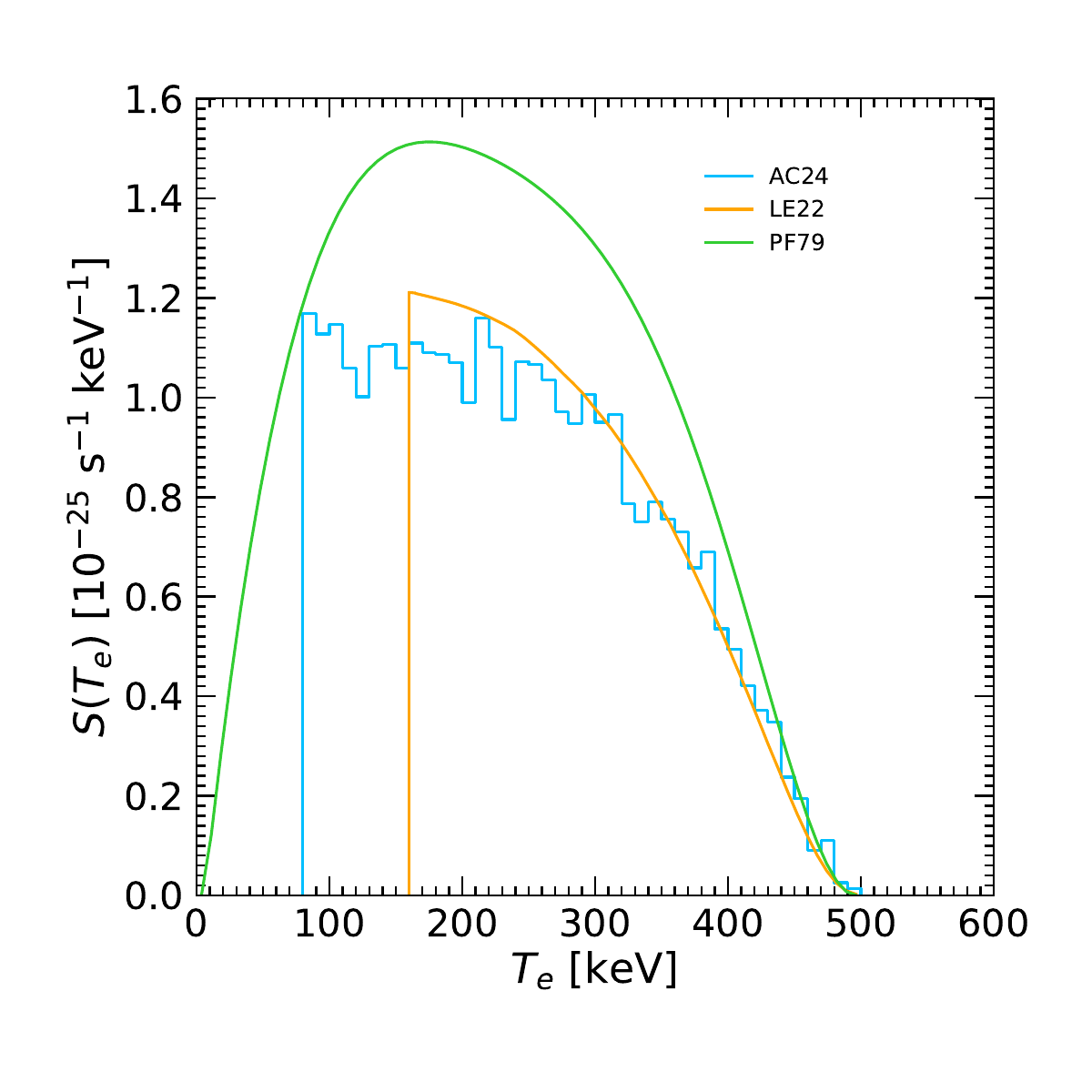}
\vspace*{-8mm}
\caption{\label{Fig_01}
\footnotesize \color{black}
$^{115}$In decay: Input energy spectra corresponding to the three independent datasets 
AC24 \cite{Pagnanini:2024qmi},
LE22 \cite{Leder:2022beq}, and
PF79 \cite{Pfeiffer:1979zz}, as used in our SMM analysis. 
See the text for details.}
\end{minipage}
\end{figure}

Figure~\ref{Fig_01} shows the input energy spectra in units of decays/s/keV. By eye, the spectra show relative differences in normalization and shape, that will be compared and discussed at the end of this Section. By means of Eqs.~(\ref{moment0}) and (\ref{moment}), the spectra in Fig.~\ref{Fig_01}
are converted into experimental moments $\mu^e_n$, as listed in Table~\ref{TabI}. 
As we did in \cite{Kostensalo:2023xzu} for $^{113}$Cd, we have verified also for
$^{115}$In that, for any practical purpose, there is no need to go beyond $N=6$ moments; actually,
significant information can be derived just from $\mu_0$ (spectrum area above threshold) and $\mu_1$ (average energy).  

A final remark is in order. We are unable to quantify also the uncertainties to be attached
to the spectra in Fig.~\ref{Fig_01}, that
depend on a number of
unpublished details about background modeling and subtraction, instrumental effects, 
and systematic errors. This detailed information is included in full-fledged data fits by the experimental collaborations
but is generally not accessible to external users. 
In any case, as shown in \cite{Kostensalo:2023xzu} for $^{113}$Cd  and discussed below for $^{115}$In, 
we can grasp various published results and reveal further implications of the decay spectra,
by applying the SMM to the limited information displayed in Fig.~\ref{Fig_01} and 
summarized in Table~\ref{TabI}, with no attempt to evaluate uncertainties for the spectra and their moments.

\begin{table}[b!]
\centering
\resizebox{.9\textwidth}{!}{\begin{minipage}{\textwidth}
\caption{\label{TabI} 
For each of the three datasets in Fig.~\ref{Fig_01}, the energy threshold and the moments
 $\mu^e_n$ are reported (up to $n=6$). The last two columns report the decay half-life $t_{1/2}$ as quoted 
by the experimental collaborations in the corresponding reference.
}
\begin{ruledtabular}
\begin{tabular}{ccccccccccccc}
	 & $T_\mathrm{thr}$ & $\mu^e_0$ & $\mu^e_1$ & $\mu^e_2$ & $\mu^e_3$ & $\mu^e_4$ & $\mu^e_5$  & $\mu^e_6$ 
&$t_{1/2}$&  \\
Dataset	 & [keV] & $[10^{-23}\,\mathrm{s}^{-1}]$ & $[10^2\,\mathrm{keV}]$ & $[10^4\,\mathrm{keV^2}]$ 
 & $[10^7\,\mathrm{keV^3}]$ & $[10^9\,\mathrm{keV^4}]$ & $[10^{12}\,\mathrm{keV^5}]$  & $[10^{14}\,\mathrm{keV^6}]$ 
 &$[10^{14}$~y]& Ref. \\
\hline
  AC24 	& 80  & $3.18$ & $2.43$ & $6.89$ & $2.17$ & $7.41$ & $2.66$ & $9.96$ & 5.26 & \cite{Pagnanini:2024qmi}\\		
  LE22 	& 160 & $2.54$ & $2.79$ & $8.37$ & $2.68$ & $9.08$ & $3.22$ & $11.9$ & 5.18 & \cite{Leder:2022beq}\\
  PF79 	& 0   & $4.98$ & $2.22$ & $6.08$ & $1.88$ & $6.31$ & $2.23$ & $8.19$ & 4.41 & \cite{Pfeiffer:1979zz}\\		
\end{tabular}
\end{ruledtabular}
\end{minipage}}
\end{table}

\vspace*{-3mm}
\subsection{AC24 input dataset}
\label{Sec:AC24}

Within the project ``Array of Cryogenic Calorimeters to Evaluate Spectral Shape'' (ACCESS) \cite{ACCESS:2023gdy},
a measurement of the $^{115}$In $\beta$ decay was performed by using 1.91~g of Indium iodide crystals and cryogenic calorimeters. The energy spectrum results were reported in \cite{Pagnanini:2024qmi} 
above a threshold $T_\mathrm{thr}=80$~keV, at the best-fit point for the sum of signal plus background {\color{black} (}see Fig.~2 therein{\color{black})}. The main input  corresponds to
the background-subtracted signal histogram in that figure. We adopt a digitized version of 
such histogram, as well as an efficiency-corrected value of 19739 decays/mol/s for the $^{115}$In
signal rate above threshold \cite{PrivatePagnanini}, fixing the zeroth moment as
$\mu^e_0=3.28\times 10^{-23}$~s$^{-1}$ in our notation. The resulting input spectrum, named 
AC24 dataset, is shown in Fig.~\ref{Fig_01} as a light blue histogram. The shape of the spectrum
determines the higher moments $\mu^e_n$ ($n\geq 1$), as listed in Table~\ref{TabI}.

For completeness, in Table~\ref{TabI}
we also report the $^{115}$In $\beta$ decay half-life value  $t_{1/2}=5.26\times 10^{14}$~y estimated in \cite{Pagnanini:2024qmi}. 
{\color{black}As already remarked, we do not take $t_{1/2}$ as an experimental input,
since it depends on the theoretical assumptions underlying the extrapolation below the measurement threshold.} 

\vspace*{-3mm}
\subsection{LE22 input dataset}
\label{Sec:LE22}
\vspace*{-1mm}

The $^{115}$In $\beta$ decay was measured in \cite{Leder:2022beq}
with 10.3~g of LiInSe${}_2$ crystals, acting 
both as a source and a high-resolution bolometric detector. The signal includes resolved $^{115}$In events 
and an unresolved pile-up component. 
The energy spectrum results were reported in \cite{Leder:2022beq}  at the best-fit {\color{black} point} for the sum of signal plus background, and above a threshold $T_\mathrm{thr}=160$~keV {\color{black} (}see Fig.~3 therein{\color{black})}. Our main
input corresponds, in the same figure, to the $^{115}$In spectrum plus the contribution of pile-up decay events ($\sim 5.5\%$ of the total events). Actually, we adopt a smoothed version of the published $^{115}$In binned spectrum 
\cite{PrivateLeder}, that we renormalize by a factor 1.055 to 
roughly account for pile-up effects, resulting in an overall decay event rate $\mu^e_0=2.54\times 10^{-23}$~s$^{-1}$ above threshold (our estimate). A more proper inclusion of pile-up events and of their effects on the decay rate can be 
performed only by the experimental
collaboration through their data analysis procedure. 

The resulting input spectrum, named
LE22 dataset, is shown in Fig.~\ref{Fig_01} as an orange curve. We also report in Table~\ref{TabI} (but do not use as input {\color{black} for reasons stated in Sec.~\ref{Sec:AC24}})
the half-life $t_{1/2}=5.18\times 10^{14}$~y quoted in \cite{Leder:2022beq}. 

\vspace*{-3mm}
\subsection{PF79 input dataset}
\label{Sec:PF79}
\vspace*{-1mm}

An older measurement of the $^{115}$In spectrum was performed in 
\cite{Pfeiffer:1979zz} using an Indium-loaded scintillation detector. For our purposes, the main results of that work are summarized by a polynomial fit reconstruction of the 
decay spectrum {\color{black} after background subtraction}, as shown in Fig.~8 therein (dotted curve). A few comments are in order.
On the one hand, the endpoint $Q=468.8$~keV estimated in \cite{Pfeiffer:1979zz} is 6\% lower 
than the current value \cite{Blachot:2012tby}, indicating possible energy-scale systematics. We simply assume that 
the $Q$-value mismatch can be corrected by an overall {\color{black} energy} 
rescaling, $T_e\to 1.06\times T_e$. On the other hand, the
energy threshold is not explicitly 
mentioned in \cite{Pfeiffer:1979zz}, although it appears to be rather low ($50$~keV or less, from the figures), with 
the polynomial spectrum almost vanishing for $T_e\to 0$. We note that
the only normalization information available in \cite{Pfeiffer:1979zz} is the half-life, 
$t_{1/2}=4.41\times 10^{14}$~y, that is well-defined at null threshold.  

In the absence of further information,  
we adopt as our input the polynomial spectrum reconstruction of \cite{Pfeiffer:1979zz}  with rescaled energy,
and assume in practice a vanishing threshold $T_\mathrm{thr}=0$, so as to fix the zeroth moment 
($\mu^e_0=\ln 2/t_{1/2}=4.98\times 10^{-23}$~s). This is the only dataset where
the measured half-life represents an input{\color{black}, as no theory-dependent extrapolation of the low-energy spectrum is required}. The adopted input spectrum, dubbed as
PF79, is shown in Fig.~\ref{Fig_01} as a green curve. 
We have verified \emph{a posteriori} that our PF79 results 
do not significantly change, by choosing 
nonzero thresholds up to $\sim 50$--80~keV, and adjusting $\mu^e_0$ to account for the spectrum area above threshold.

\vspace*{-3mm}
\subsection{Qualitative comparison of datasets}
\label{Compare}
\vspace*{-1mm}

The three spectra in Fig.~\ref{Fig_01}  represent independent measurements of the same $^{115}$In $\beta$ absolute decay spectrum, and in principle they should be quite similar to each other in both normalization and shape
(above threshold). In practice,
qualitative differences emerge already at visual inspection, as discussed below. Quantitative implications 
will be derived in Sec.~\ref{Sec:Comparison}.

At a glance, the LE22 and AC24 spectra in Fig.~\ref{Fig_01} appear to be in qualitative
agreement, although 
the LE22 spectrum tends to overshoot the AC24 one at low energy, suggesting a higher decay rate of
the former with respect to the latter. Modulo different sub-threshold shapes, this
expectation is consistent with the smaller $t_{1/2}$ value quoted by LE22 with respect to AC24.
However, just by eye, the LE22 and AC24 spectra do not allow to {\color{black} reliably predict whether} their sub-threshold behavior
tend to be decreasing, stationary, or increasing as $T_e\to 0$.

The PF79 data entail a decay rate noticeably higher than LE22 or AC24 at any energy.
In addition, the PF79 spectrum displays a seemingly identified peak between 150 and 200~keV (not evident
in AC24 or LE22 data), plus a rapid drop at low energy (not emerging in AC24 data).
In terms of total decay rate (area
under the spectrum), the high 
PF79 normalization is only partly compensated by it low-energy decrease, consistent with
a quoted PF79 half-life significantly lower (roughly by $-15\%$) than those quoted by AC24  and LE22.
We emphasize that comparing half-lives alone mixes two different 
issues (overall normalization and low-energy shapes) that should be considered separately in each dataset, 
as correctly implemented in the spectral moment method.

\begin{figure}[t]
\begin{minipage}[c]{0.7\textwidth}
\includegraphics[width=0.72\textwidth]{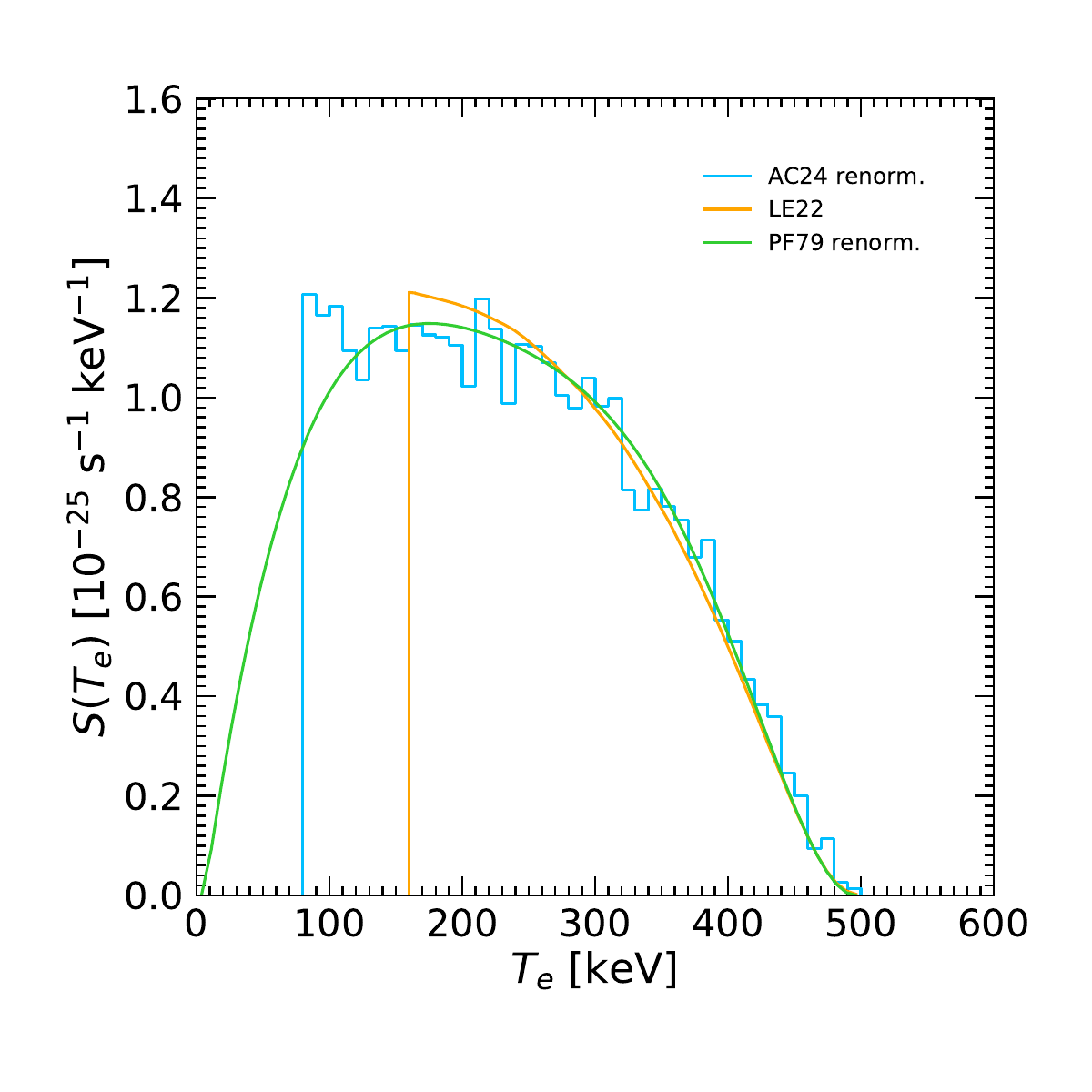}
\vspace*{-8mm}
\caption{\label{Fig_02}
\footnotesize \color{black} Energy spectra as in Fig.~\ref{Fig_01}, but with AC24 and PF79 
renormalized by factors 1.033 and 0.759, respectively, so as to cover the same area as LE22 above
160~keV.} 
\end{minipage}
\end{figure}

{\color{black}Independent of} any assumption about the low-energy
behaviour, a shape-only comparison of the three spectra 
can be performed above the highest threshold, corresponding to $T_\mathrm{thr}=160$~keV for
the LE22 dataset. Figure~\ref{Fig_02} shows the comparison of the three datasets, where  LE22 is
the same as in Fig.~\ref{Fig_01}, while
AC24 and PF79 are renormalized by factors 1.033 and 0.759, respectively, 
in order to have the same area as LE22 above threshold. 
The small normalization adjustment  
between AC24 and LE22 corroborates the noted compatibility of the two spectra, 
in contrast with their large normalization difference with PF79. 
Concerning the spectral shapes, the three renormalized spectra are in reasonable agreement  
above 160 keV, while below this value the PF79 spectrum shows a low-energy decrease not supported by AC24.

{\color{black}In summary}, visual inspection shows 
relative differences in normalization and shape among the three input spectra, 
that might be due to possible experimental systematics. 
Normalization differences are relatively small
between the AC24 and LE22 datasets, but they are relatively large between these data and the {\color{black}much older} PF79 dataset.
Shape differences among the three datasets are relatively small at high energy, but appear to
increase at low energy. 
Both small and large differences will affect the comparison of data with theoretical calculations,
the evaluation of their free parameters, and the implications for sub-threshold shapes and decay half-lives.

\vspace*{-2mm}
\section{$^{115}\mathrm{\bf In}$ spectrum: Theoretical calculations and free parameters}
\label{Sec:Models}

Concerning the calculation of the 
theoretical $^{115}$In $\beta$-decay spectrum $S^t(T_e)$, we adopt the general approach discussed in 
\cite{Haaranen:2017ovc}, to which we refer the reader for details and bibliography. Here we remind that, apart from overall factors, $S^t(T_e)$ is a sum over squares and products of linear combinations of NME, arising from a multipole expansion of the nuclear current in the formalism of Behrens and B\"uhring \cite{Behrens1982}. The power-series expansion results in an expression including both vector NME ${}^V\!{\cal{M}}^{(m)}_{\rm KLs}$ (weighted by $g_{\rm V}$) and axial-vector NME {\color{black}${}^A\!{\cal{M}}^{(m)}_{\rm KLs}$} (weighted by $g_{\rm A}$). As a consequence, by defining the ratio
\begin{equation}
\label{ratio}
r = g_{\rm A}/g_{\rm V}
\end{equation}
the quadratic-form structure of $S^t(T_e)$ leads to the general expression \cite{Haaranen:2016rzs,Kostensalo:2023xzu},
\begin{equation}
\label{VA}
S^t = g_{\rm V}^2(S^t_{\rm V}+r S^t_{\rm VA}+r^2 S^t_{\rm A})\ ,
\end{equation}
where the three spectrum terms contain vector-only (V), axial-only (A), and mixed vector and axial-vector (VA) NME contributions.
As in \cite{Kostensalo:2023xzu}, in accordance with the conserved vector current (CVC) hypothesis, we shall assume 
that $g_{\rm V}=1$, while we shall treat $r$ (i.e., $g_{\rm A})$ as a free parameter.  Following \cite{Haaranen:2017ovc},
we consider three nuclear models for quantitative calculation of the NME: the microscopic interacting boson-fermion model (IBFM-2), the microscopic quasiparticle-phonon model (MQPM) and the interacting shell model (ISM).

For fourth-forbidden unique decays there are four leading-order NME, with the dominant matrix elements being ${}^V\!{\cal{M}}^{(0)}_{440}$ and ${}^A{\cal{M}}^{(0)}_{441}$, and with significantly smaller contributions coming from ${}^V\!{\cal{M}}^{(0)}_{431}$ and ${}^A\!{\cal{M}}^{(0)}_{541}$. The power expansion at next-to-leading order results in a total of 
45 NME \cite{Haaranen:2017ovc} that depend on $T_e$ and on powers of the nuclear radius. 
As discussed in detail in \cite{Haaranen:2016rzs,Haaranen:2017ovc}, for $r\sim 1$ the quadratic form in Eq.~(\ref{VA}) entails delicate cancellations among large NME, that lead to large spectral variations in normalization and shape{\color{black}. While the experimental spectrum and half-life may be reproduced for some value of $r$, it is not guarantied that there is any value of $r$ which could reproduce both the half-life and spectrum simultaneously.}
In this context, a small NME may play a 
relevant role by modulating the residuals of large NME cancellations, especially if its numerical value is
rather uncertain from a nuclear model viewpoint. It was realized in 
\cite{Kirsebom:2019tjd,Kumar:2020vxr,Kostensalo:2020gha}
 that this role can be played by
so-called small relativistic NME, dubbed as $s$-NME in \cite{Kostensalo:2020gha} and just as $s$ 
in \cite{Kostensalo:2023xzu} and herein, where
(in units of fm$^3$):
\begin{equation}
\label{sNME}
s = {}^V\!{\cal{M}}^{(0)}_{431}\ .
\end{equation}

By using both $r$ and $s$ as free parameters in detailed analyses of experimental
data, simultaneous fits to both the measured spectrum shape and the inferred half-life were achieved 
for $^{113}$Cd decay in \cite{Kostensalo:2020gha}
and for $^{115}$In decay in \cite{Leder:2022beq}, allowing estimates of $g_{\rm A}$-quenching effects in the two nuclei.
For $^{113}$Cd decay, results similar to those in \cite{Kostensalo:2020gha}  were also obtained by applying 
the spectral moment method \cite{Kostensalo:2023xzu} to an independent dataset \cite{Belli:2007zza}. 
In general, the SMM offers a simple and useful guidance in the $(r,\,s)$ parameter space, 
independently of the  specific nucleus or dataset, as
briefly described below (see \cite{Kostensalo:2023xzu} for further details).

Since the $(r,\,s)$ parameters enter in the spectrum $S^t(T_e)$ up to their second power, any theoretical spectral moment $\mu^t_{n}$  is related to quadratic forms in $(r,\,s)$ via integrals of the kind 
[see Eqs.~(\ref{moment0}) and (\ref{moment})]: 
\begin{equation}
\label{quadra}
\int_{T_{\rm thr}}^{Q} S^t(T_e)\,T_e^n\,dT_e = \sum_{i+j\leq 2}a^n_{ij}\, r^i\, s^j \ ,
\end{equation}
where the coefficients $a^n_{ij}$ (with $n$ being a superscript, not a power) are calculable numbers within each 
adopted nuclear model. The $0^{\rm th}$ moment $\mu^t_0(r,\,s)$ corresponds to the above quadratic form with $n=0$
[Eq.~(\ref{moment0})],
while the  $n^{\rm th}$ moment $\mu^t_n(r,\,s)$ for $n\geq 1$ corresponds to a ratio of quadratic forms, with 
index $n$ at numerator and zero at denominator [Eq.~(\ref{moment})]. 
Thus, by equating the theoretical and experimental moments,
\begin{equation}
\label{impose}
\mu^t_n(r,\,s) = \mu^e_n \ (n\geq 0)\ ,
\end{equation}
one gets, geometrically, conic sections in the  $(r,\,s)$ plane. In particular, it turns out that for $n=0$ the
conic section is an ellipse, while for $n\geq 1$ the conic sections form a bundle of hyperbolas \cite{Kostensalo:2023xzu}.

In principle, perfect agreement between theory and data would correspond to a unique intersection point between the
ellipse and all the hyperbolas (with sparse intersections around a secondary location) \cite{Kostensalo:2023xzu}. At that unique point, the theoretical and experimental spectra would coincide in both normalization ($\mu_0$) and shape ($\mu_n$ with $n\geq 1$).
In practice, since theory and data never match perfectly, each hyperbola intersects the ellipse in different points. The smaller (larger) spread of such points indicates qualitatively a better (worse) agreement between theory and data. Useful guidance about the preferred $(r,\,s)$ parameter
values can thus be simply obtained by looking at the intersections of the $\mu_0$ ellipse with the $\mu_1$ 
hyperbola, and at the local spread of intersections with a few higher-moment hyperbolas $\mu_{2,3,\dots,6}$ \cite{Kostensalo:2023xzu}. Note that at the intersection point(s) of $\mu_0$ and $\mu_1$, 
at least the spectrum area and 
the average $\beta$-particle energy (both defined above threshold) are matched by construction.

In the next Section, the SMM approach will be applied to the
three independent $^{115}$In decay datasets previously named as AC24 \cite{Pagnanini:2024qmi}, LE22 \cite{Leder:2022beq}
and PF79 \cite{Pfeiffer:1979zz}. We shall recover various results already obtained through refined data fits
in \cite{Pagnanini:2024qmi} and partly in \cite{Leder:2022beq}. In addition, we shall highlight 
various differences in the low-energy behavior of the spectra, that are entangled with 
$(r,\,s)$ and $t_{1/2}$ estimates.

\begin{table}[t]
\centering
\resizebox{.5\textwidth}{!}{\begin{minipage}{0.69\textwidth}
\caption{\label{TabII} 
For each dataset and nuclear model considered, we report the $(r,\,s)$ parameter values that solve the first two moment
equations, $\mu^e_0=\mu^t_0$ and $\mu^e_1=\mu^t_1$, for $s<0$ and $s>0$.  We also report the corresponding theoretical half-life $t_{1/2}$ in units of $10^{14}$~y.}
\begin{ruledtabular}
\begin{tabular}{ccccc}
Dataset & Model & $r$ & $s$ & $t_{1/2}$ [$10^{14}$~y]	\\
\hline
AC24 	&	IBFM-2	& 0.961 & $-1.125$ & 5.51  \\
		&			& 1.183 & $+1.036$ & 5.35  \\
		&	MQPM	& 0.994 & $-0.994$ & 5.62  \\ 
		&			& 1.103 & $+0.849$ & 5.12  \\
		&	ISM		& 0.888 & $-1.162$ & 5.60  \\
		&			& 0.970 & $+1.060$ & 5.24  \\
\hline
LE22	&	IBFM-2	& 0.796 & $-0.970$ & 4.93  \\
		&			&  ---  &  ---     & ---   \\
		&	MQPM	& 0.968 & $-0.910$ & 5.26  \\
		&			& 1.239 & $+0.268$ & 3.92  \\
		&	ISM		& 0.849 & $-1.056$ & 5.21  \\
		&			& 1.105 & $+0.629$ & 4.20  \\
\hline
PF79	&	IBFM-2	& 1.255 & $-1.672$ & 4.41  \\
		&			& 0.808 & $+1.769$ & 4.41  \\
		&	MQPM	& 1.077 & $-1.556$ & 4.41  \\
		&			& 0.970 & $+1.680$ & 4.41  \\
		&	ISM		& 0.987 & $-1.704$ & 4.41  \\
		&			& 0.826 & $+1.778$ & 4.41  
\end{tabular}
\end{ruledtabular}
\end{minipage}}
\end{table}

\vspace*{-2mm}
\section{Comparison of theoretical and experimental spectra}
\label{Sec:Comparison}

In this Section{\color{black},} we compare theoretical and experimental spectra for $^{115}$In decay via the SMM. We compute theoretical spectra $S^t(T_e)$ as functions of the free parameters $(r,\,s)$ in each of the three reference nuclear models (IBFM-2, MQPM and ISM), adopting the one-body transitions densities 
reported in \cite{Haaranen:2017ovc}. We consider the three experimental spectra $S^e(T_e)$ for the AC24, LE22, and PF79 
datasets as reported in Fig.~\ref{Fig_01}, together with their nominal thresholds set at $T_\mathrm{thr}=80$, 160 and 0~keV, respectively. For each pair $(r,\,s)$, we compute the theoretical moments $\mu^t_n(r,\,s)$, and 
equate them [via Eq.~(\ref{impose})] to the corresponding experimental moments $\mu^e_n$ reported in Table~\ref{TabI}. 

In particular, by solving Eq.~(\ref{impose}) for the first two moments ($n=0,\,1$), we get the $(r,\,s)$ solutions reported in Table~\ref{TabII}. In the same Table we also report, as a by-product,
the theoretically estimated $^{115}$In decay half-life $t_{1/2}$. As already noted, the zeroth moment and the half-life contain equivalent information only for a null threshold [via Eq.~(\ref{rate})], so that the theoretical value of
$t_{1/2}$ is uniquely fixed by $\mu_0^e$ for PF79, while it depends on $(r,\,s)$ for AC24 and LE22.
Note that in Table~\ref{TabII}, for each dataset, there are generally two degenerate $(r,\,s)$ solutions for negative and positive values of $s$, corresponding to different theoretical spectra. The degeneracy may be lifted by looking at the spread of moment intersections: the smaller the spread, the better the match between theory and data, 
as discussed below.

\vspace*{-3mm}
\subsection{Comparison with AC24 data}
\label{Sec:ComparisonAC24}
\vspace*{-1mm}

\begin{figure}[b!]
\begin{minipage}[c]{\textwidth}
\vspace*{-4mm}
\includegraphics[width=0.9\textwidth]{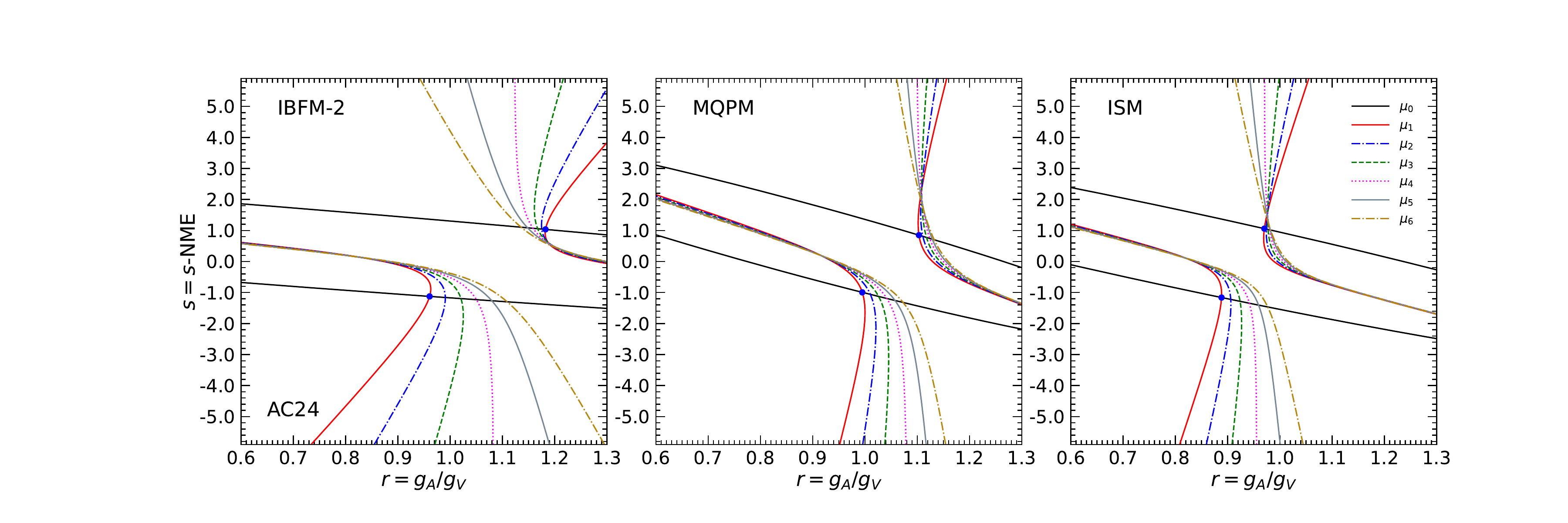}
\vspace*{-4mm}
\caption{\label{Fig_03}
\footnotesize Analysis of AC24 dataset in the plane charted by the free parameters $r=g_{\rm A}/g_{\rm V}$
and $s=s$-NME, with 
isolines corresponding to the equality of theoretical and experimental moments 
$\mu^t_n=\mu^e_n$ (up to $n=6$). The left, middle and right panels correspond to the IBFM-2, MQPM{\color{black},} and ISM model, respectively.
In each panel, the slanted ellipse is determined by $\mu_0$, while the bundle of hyperbolas 
is determined by $\mu_1$ and higher moments. The intersections of $\mu_0$ and $\mu_1$ isolines are
denoted by dots. The local spread of intersections of $\mu_0$ with higher moments would vanish
for a perfect match between theoretical and experimental spectra. See the text for details.
} \end{minipage}
\end{figure}

Figure~\ref{Fig_03} shows the loci of points in the $(r,\,s)$ plane fulfilling Eq.~(\ref{impose})
up to $n=6$, using the experimental moments $\mu_n^e$ from the AC24 dataset as reported in Table~\ref{TabI}.
The left, middle and right panels correspond to the IBFM-2, MQPM{\color{black},} and ISM model, respectively.
In each panel, the $\mu_n$ isolines correspond to conic sections \cite{Kostensalo:2023xzu}: 
the elongated, slanted ellipse is determined by $\mu_0$, while the bundle of hyperbolas is 
determined by $\mu_1$ and higher moments. The intersections of $\mu_0$ and $\mu_1$ isolines are
denoted by two dots, identifying the $(r,\,s)$ pair at positive and negative values of $s$ reported
in Table~\ref{TabII}. In all cases, these dots correspond to quenched values of $g_{\rm A}$ ($r<1.26$)
and to values  $|s|\simeq O(1)$.

In the ideal case of perfect match between theory and data, all isolines 
should intersect in a single point, leading to a unique $(r,\,s)$ solution; conversely, any mismatch
between theory and data would lead to somewhat different intersections of $\mu_0$ and $\mu_n$ ($n\geq 2$) isolines. The local spread of such intersections provides a visual appreciation of the (dis)agreement between 
nuclear model calculation and AC24 data. In each panel, 
the solution (dot) for $s>0$ entails a smaller spread than for $s<0$ and is thus in better agreement with the data. 
For all models, the $s>0$ solutions correspond to moderate quenching ($r\simeq 1.0$--1.2), as compared to 
the $s<0$ solutions ($r\simeq 0.9$--1).

Our spectral moment method results for AC24 data are in striking agreement with those obtained in 
\cite{Pagnanini:2024qmi} through a refined fit to signal plus background, using the enhanced spectral shape method and the same nuclear models. In particular, the six $(r,\,s)$ pairs 
and $t_{1/2}$ values reported in our Table~\ref{TabII} for AC24 agree within few percent with the best-fit results reported 
in Table~II of \cite{Pagnanini:2024qmi} (see the ``matched half life'' entries therein), with a consistent preference 
for $s>0$ (``positive solutions'' therein). We also get a clear understanding---in terms of our moment isolines---of the 
graphical fit results reported in Fig.~3 of \cite{Pagnanini:2024qmi}:
in that figure, the fit to $t_{1/2}$ leads to an elliptical allowed band (akin to our $\mu_0$ isoline), while the 
pixels sampling the ``shape-only'' fit 
{\color{black} are scattered around two curved branches} (akin to our bundle of $\mu_n$ isolines for $n\geq 1$).

The very good correspondence between our results and those in \cite{Pagnanini:2024qmi} shows that our SMM can capture
the preferred $(r,\,s)$ values in a simple and intuitive way. However, the correspondence can never be exact, for two different reasons. On the one hand, our method has some limitations, since we cannot evaluate parameter errors, 
as it is possible in refined data analyses by the experimental collaborations. 
On the other hand, our SMM approach has the advantage of using
only observable information above threshold, with no sub-threshold information as embedded in $t_{1/2}$ (a fitted
parameter in the SSM approach). Therefore, we are able to present an unbiased discussion of the low-energy behavior of the theoretical spectra, constrained only by experimental data above threshold and with no prior assumption.

\begin{figure}[t!]
\begin{minipage}[c]{0.7\textwidth}
\includegraphics[width=0.72\textwidth]{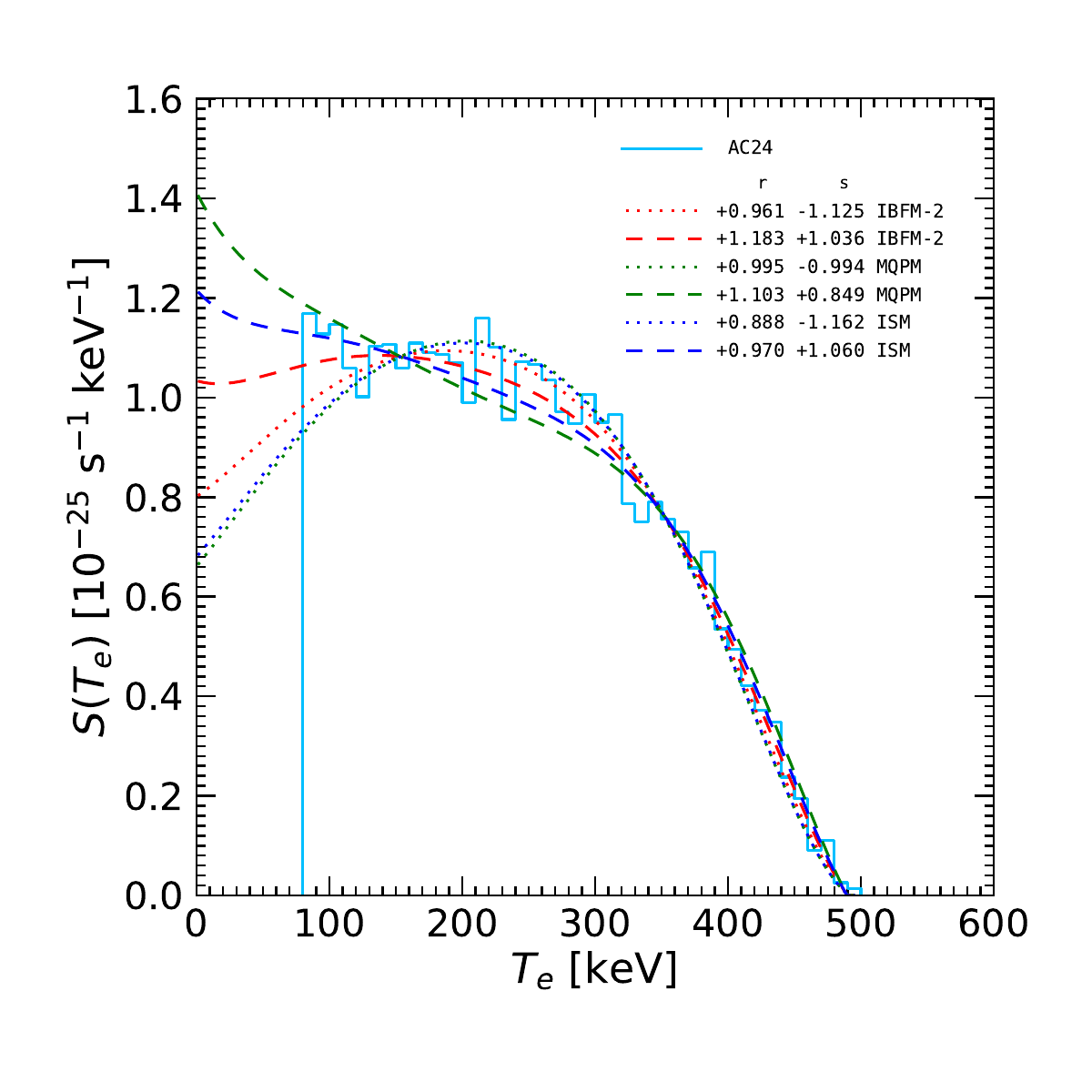}
\vspace*{-8mm}
\caption{\label{Fig_04}
\footnotesize Comparison of the AC24 dataset with the nuclear model spectra, calculated for the
same $(r,\,s)$ values as in Table~\ref{TabII}. The relative spectral differences appear to be 
moderate at intermediate and high energies, and more significant at low (and sub-threshold) energies.
The $s>0$ cases are compatible with both decreasing and increasing 
spectral shapes below threshold.  
} \end{minipage}
\end{figure}

Figure~\ref{Fig_04} shows the six theoretical spectra corresponding to  the six AC24 entries 
in Table~\ref{TabII}, namely, to the six $(r,\,s)$ solutions (dots) in Fig.~\ref{Fig_03}, compared
to our input AC24 dataset. The spectra with {\color{black} $s<0$} show clear peaks, 
{\color{black} that tend to slightly overshoot the data at intermediate energies, 
and to rapidly decrease below threshold.} The (preferred)
spectra at $s>0$ instead have either a moderate peak with a slight low-energy decrease (IBFM-2), or no peak at all
with a noticeable low-energy increase (ISM and especially MQPM). These variants affect the
(unobserved) area below threshold, that contributes to the total estimated event rate and thus 
to the inferred value of $t_{1/2}$. As far as the low-energy spectrum remains ambiguous,
the estimated value of $t_{1/2}$  also remains {\color{black} somewhat} uncertain.

Assuming the best-fit results from \cite{Pagnanini:2024qmi} (obtained for IBFM-2 at $r\simeq 1.2$ and $s\simeq 1$), the expected
spectrum would correspond to the red, long-dashed curve in Figure~\ref{Fig_04}, characterized by a slight
low-energy decrease.  Further spectral data with higher
statistics and lower threshold appears necessary to disentangle this best-fit prediction from its closest model
variants (MQPM and ISM at $s>0$), characterized by a low-energy increase.

\vspace*{-3mm}
\subsection{Comparison with LE22 data}
\label{Sec:ComparisonLE22}
\vspace*{-1mm}

\begin{figure}[t!]
\begin{minipage}[c]{\textwidth}
\includegraphics[width=0.9\textwidth]{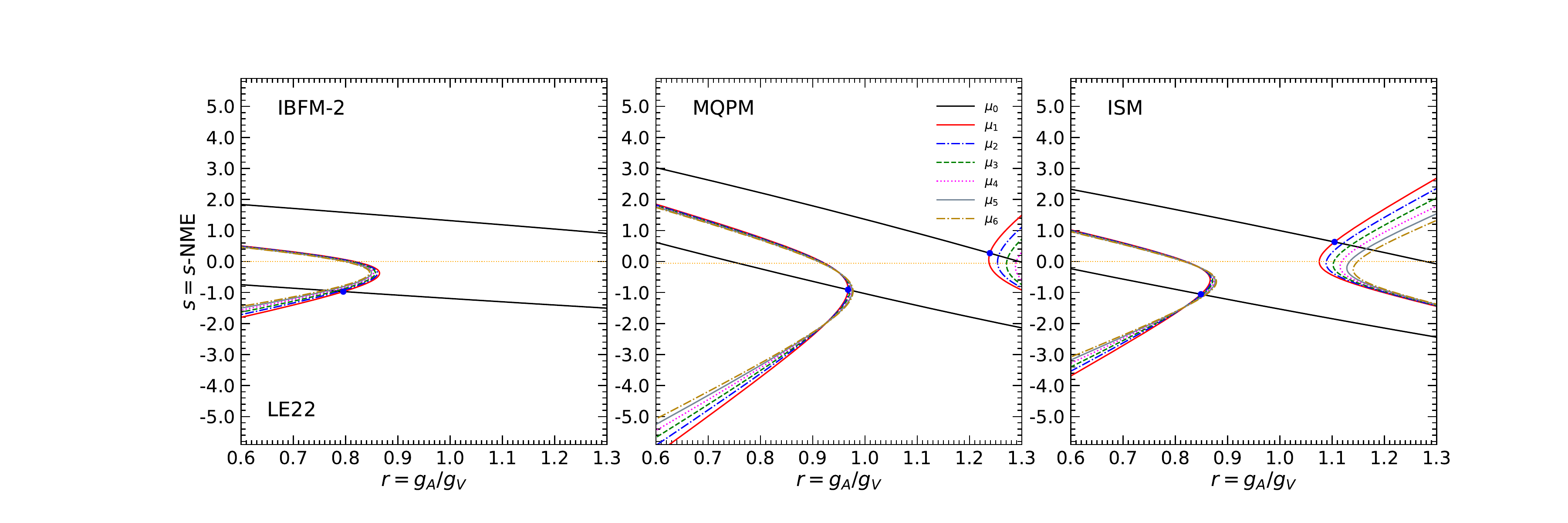}
\vspace*{-4mm}
\caption{\label{Fig_05}
\footnotesize As in Fig.~\ref{Fig_03}, but for the LE22 data analysis. Note the absence of a second solution
in the left panel. The horizontal dotted lines correspond to $s$ values fixed by models. See the text
for details.
} \end{minipage}
\end{figure}

Figure~\ref{Fig_05} is constructed as Fig.~\ref{Fig_03} but using LE22 data. For the IBFM-2 model, one branch of the
bundle of hyperbolas lies outside the considered parameter range for $r$, so that only five solutions (dots) out of 
six are shown in the figure and reported in Table~\ref{TabII}. (The unreported solution is also disfavored by
a rather large spread of intersections, not shown).  
From the reduced spread of local intersection points, 
a preference emerges for solutions at $s<0$, 
characterized by significant quenching, $r\simeq 0.8$--1 (depending on the model).
These results for LE22 data are somewhat different from those obtained for AC24 data, despite the fact the 
the two datasets appear to be very similar at high energy and moderately different just above the LE22 threshold
(see Fig.~\ref{Fig_01}). This fact suggests that the determination of the $r$ and $s$ parameters is 
quite sensitive to detailed spectral features, especially at relatively low energies.

\begin{figure}[b!]
\begin{minipage}[c]{0.7\textwidth}
\includegraphics[width=0.72\textwidth]{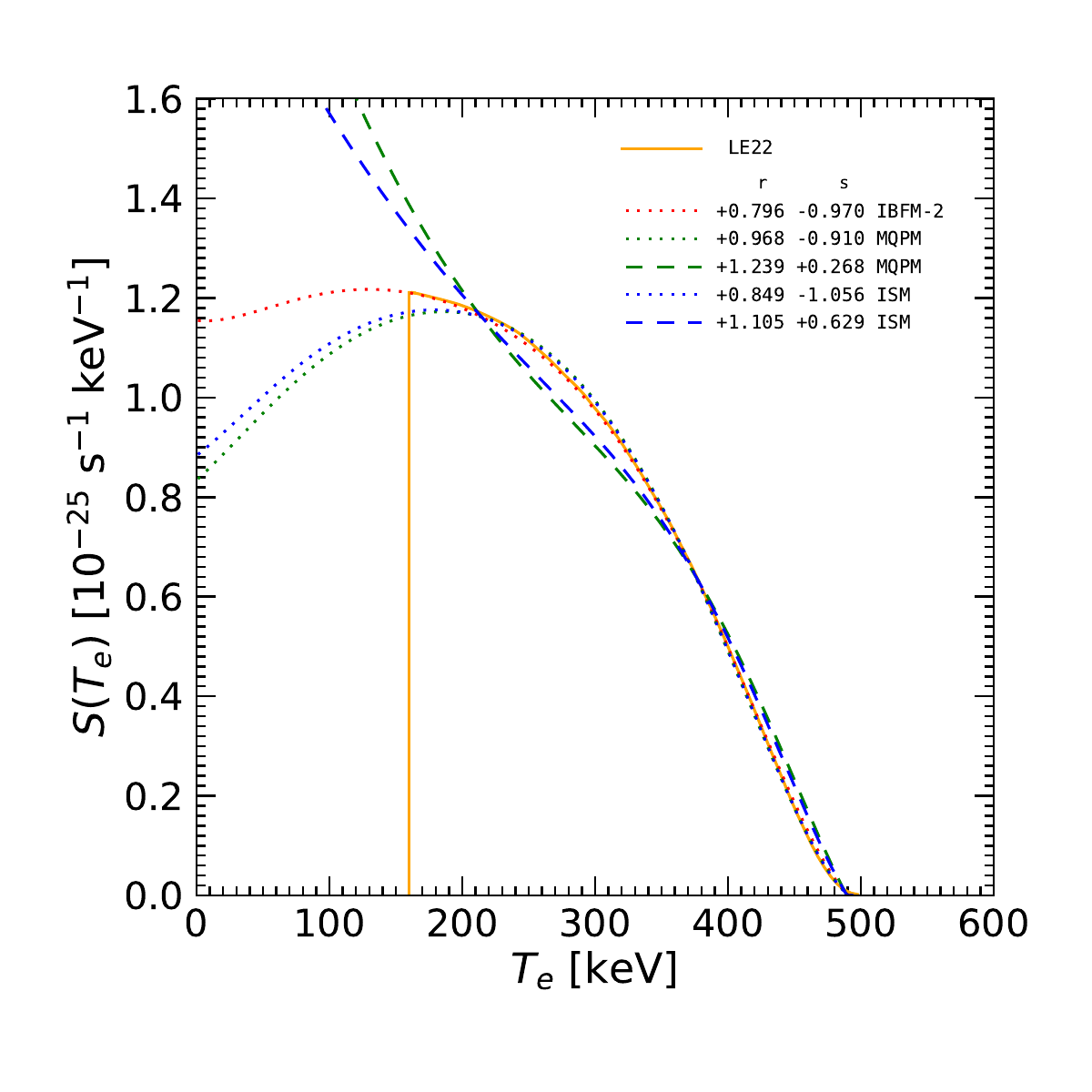}
\vspace*{-8mm}
\caption{\label{Fig_06}
\footnotesize As in Fig.~\ref{Fig_04}, but for the LE22 data analysis.
} \end{minipage}
\end{figure}

This hint is confirmed by the results shown in Fig.~\ref{Fig_06}, 
that is analogous Fig.~\ref{Fig_04} but using LE22 data. The worse solutions (at $s>0$) 
show shape deviations and a rapid low-energy increase, not supported by the input data.
The preferred solutions at $s<0$ for the three models
reproduce well the LE22 data, but predict a low-energy decrease of the spectrum, generally
in contrast with the preferred  solutions for AC24 at $s>0$. Only the IBFM-2 model predicts 
a slight low-energy decrease of the spectrum both for LE22 (at $s<0$) and for AC24 (at $s>0$), but with 
a normalization different by $\sim 10\%$ as $T_e\to 0$ (compare  Figs.~\ref{Fig_06} and \ref{Fig_04}).
Once more, we surmise that further data with the smallest possible threshold would be very useful to 
solve residual ambiguities, affecting the low-energy shape and the overall normalization of 
$^{115}$In spectra.  A theoretical assessment of sign$(s)$ in the various models would also be beneficial.

The above LE22 results, obtained through the SMM by assuming
two free parameters $(r,\,s)$,
are not immediately comparable with those obtained in \cite{Leder:2022beq}. In \cite{Leder:2022beq},  
a shape-only analysis was performed with free $r$, while $s$ was fixed
by the nuclear models at face value. In particular, for $^{115}$In, the IBFM-2 and ISM model 
provide the value $s=0$ (due to limitations of the restricted model space), while the MQPM model 
predicts a small negative value, $s=-0.054$. See also \cite{Kostensalo:2023xzu} for
an analogous discussion of model $s$-values for $^{113}$Cd, and \cite{Haaranen:2017ovc} for 
more general considerations. 

\begin{figure}[t!]
\begin{minipage}[c]{\textwidth}
\includegraphics[width=0.9\textwidth]{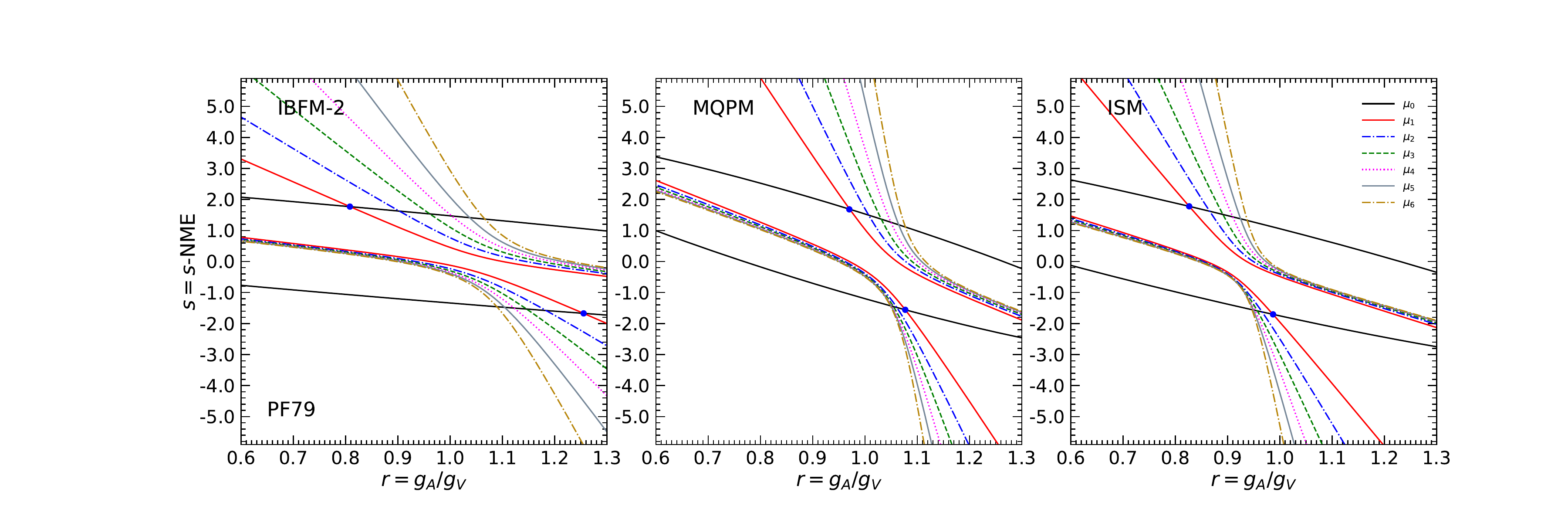}
\vspace*{-4mm}
\caption{\label{Fig_07}
\footnotesize As in Fig.~\ref{Fig_03}, but for the PF79 data analysis. 
} \end{minipage}
\end{figure}

In order to perform a comparison with \cite{Leder:2022beq}, in Fig.~\ref{Fig_05} we report  
the fixed model values of $s$ as horizontal dotted lines; their intersection with the $\mu_1$
hyperbola (with the smallest local spread of higher $\mu_n$ intersections) 
should provide an estimate of the $r$ values preferred by the shape-only analysis
of the LE22 data in \cite{Leder:2022beq}. We get $r\simeq 0.82$ (IBFM-2), 0.92 (MQPM){\color{black},} and 0.82 (ISM), in reasonable agreement 
with the corresponding best-fit values quoted in \cite{Leder:2022beq}: 0.845 (IBFM-2), 0.936 (MQPM), and 0.830 (ISM).
Figure~\ref{Fig_05} also shows that such $s$-fixed intersections are quite far from the $\mu_0$
ellipse, and thus fail to reproduce the spectrum normalization in addition to the shape,
despite the fact that the $r$ values do not change much  by taking $s$ fixed or free in any model.
At least for the LE22 dataset, the spectrum shape seems thus be more important {\color{black} than} its
normalization to determine the amount of $g_{\rm A}$ quenching. Note that LE22 
has the highest nominal threshold and thus the low-energy spectrum shape is largely unprobed.

{\color{black}In summary}, for LE22 data we can reasonably reproduce the published $r$ values at fixed $s$ \cite{Leder:2022beq}. For unconstrained $(r,\,s)$ our LE22 results are somewhat different from those
obtained above for AC24 with lower threshold. The evaluation of the free model parameters in AC24 and LE22 data 
appears to be quite sensitive to shape variants. In particular, assessing whether 
the spectra increase or decrease at low energy would provide more robust $(r,\,s)$
estimates.

\vspace*{-3mm}
\subsection{Comparison with PF79 data}
\label{Sec:ComparisonPF79}
\vspace*{-1mm}

\begin{figure}[b!]
\begin{minipage}[c]{0.7\textwidth}
\includegraphics[width=0.72\textwidth]{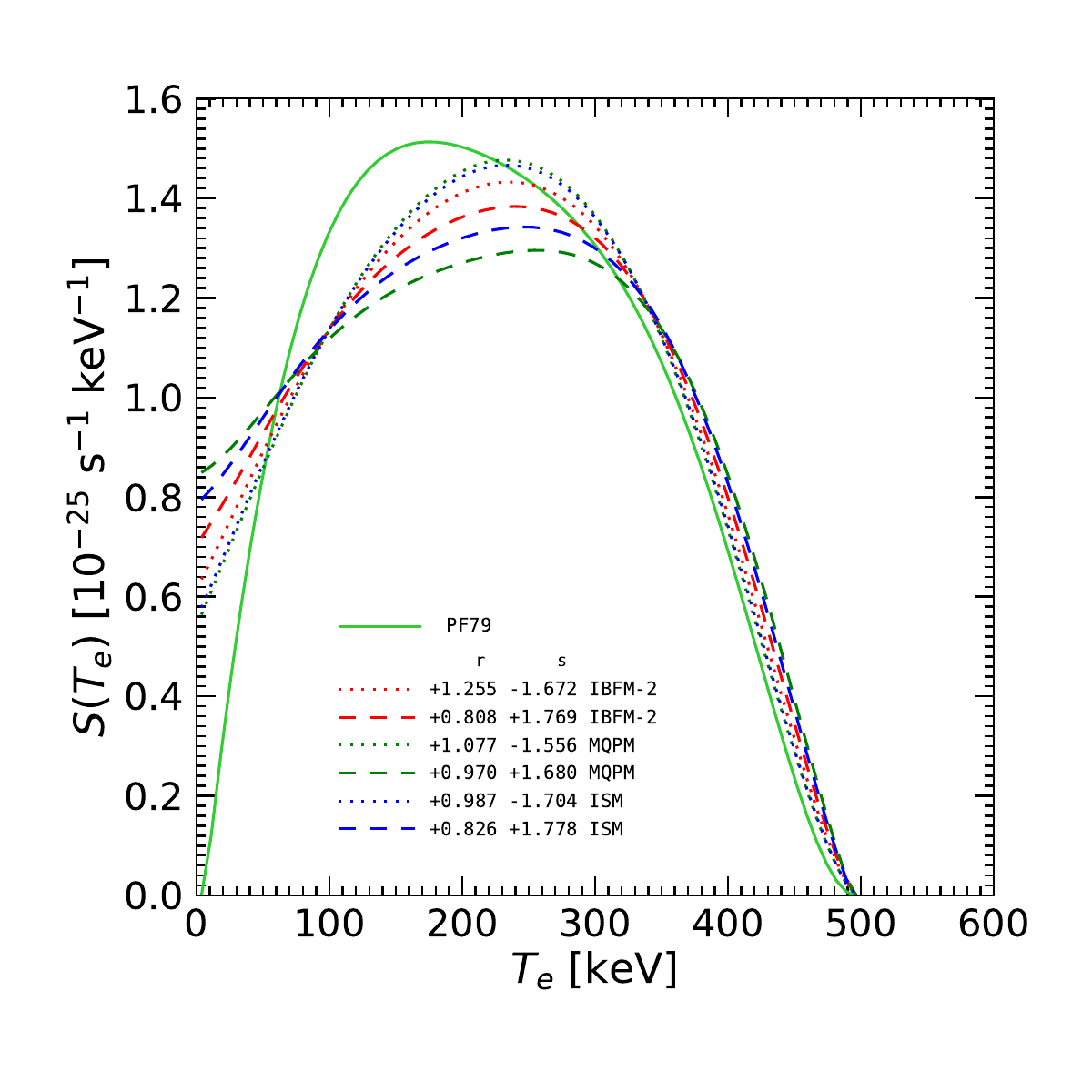}
\vspace*{-8mm}
\caption{\label{Fig_08}
\footnotesize As in Fig.~\ref{Fig_04}, but for the PF79 data analysis.
} \end{minipage}
\end{figure}

Figure~\ref{Fig_07} is analogous to Fig.~\ref{Fig_03} but using PF79 data. The spread of
intersections of the $\mu_0$ ellipse with higher-moment hyperbolas appears to be relatively large in general,
suggesting that the theoretical spectra can only roughly match the PF79 data. The smallest spread is
reached for the MQPM and ISM models at $s<0$. 

Figure~\ref{Fig_08} confirms this guess. The MQPM and ISM models at $s<0$ provide the spectra with 
the highest peak and the most rapid decrease at low energy, in qualitative agreement with PF79 input data.
However, from a quantitative viewpoint, no model is able to reproduce the peak position and 
the low-energy vanishing of the PF79 spectrum. 

The PF79 dataset shows thus peculiar aspects with respect to the AC24 and LE22 datasets,
not only in terms of model-independent differences in normalization and shape (as discussed for Fig.~\ref{Fig_01}),
but also from the viewpoint of the {\color{black} model parameter estimation}. 

Within the scope of our work, we have been able to analyze the implications of the various AC24, LE22 and PF79 
spectra in Fig.~\ref{Fig_01} but, of course, we cannot trace or guess the origin of their experimental differences in 
normalization and shape. 
As a consequence, although one can safely say that all the available data and the considered
models prefer $r<1.276$ and thus $g_{\rm A}$ quenching, more definite constraints on $r$ (as well as on the sign of
$s$) still escape precise quantification in $^{115}$In.
Constraints on the inferred half-life values are also partly uncertain, since $t_{1/2}$ is inversely
proportional to the spectrum area, which may increase or decrease together with the (poorly understood) 
low-energy spectrum shape, besides depending on the overall normalization. Further spectral data might help 
to settle these issues.

\section{Summary and perspectives}
\label{Sec:Summary}

Measuring and understanding the energy spectra of highly forbidden non-unique $\beta$ decays provides an
interesting window to neutrino and nuclear physics.
These spectra are very sensitive to the relative weight of vector and axial-vector nuclear matrix elements, and might thus
shed light on the debated issue of the effective (quenched) value of $g_{\rm A}$ in nuclear matter, that is relevant to 
$0\nu\beta\beta$ decay searches and other weak processes \cite{Suhonen:2017krv}. However, matching theoretical 
and experimental $\beta$ spectra  appears to be a very challenging
task: nuclear model results, taken at face value, may easily fail (sometimes badly) to describe either the
observed spectral shape or its absolute normalization 
(i.e., the decay rate) above the experimental threshold in $\beta$ energy.

In the context of $^{113}$Cd and $^{115}$In fourth-forbidden $\beta$ decays, 
a good theoretical description of the experimental spectrum shape (but not of its normalization) has been obtained 
in nuclear models using $g_{\rm A}$ as the only free parameter (IBFM-2, MQPM, ISM) \cite{COBRA:2018blx,Leder:2022beq} 
and, more recently, in a shell model without free parameters (RSM) \cite{DeGregorio:2024ivh}. So far, a satisfactory 
match of  
both  normalization and shape has been achieved only by using two degrees of freedom 
\cite{Kostensalo:2020gha,Pagnanini:2024qmi}, that in our work are named as $r=g_{\rm A}/g_{\rm V}$ and 
$s=s$-NME (the small and uncertain relativistic NME). 
Such nontrivial match generally requires both
$g_{\rm A}$ quenching and $|s|\sim O(1)$ in each of the examined nuclear models (IBFM-2, MQPM{\color{black},} and ISM)
as derived by detailed experimental data fits \cite{Kostensalo:2020gha,Pagnanini:2024qmi},

In a previous paper \cite{Kostensalo:2023xzu}, we have shown how to discretize and simplify the
analysis 
of the $^{113}$Cd spectrum in the $(r,\,s)$ parameter space, in terms of spectral moments $\mu_n$ defined
above the experimental threshold (spectral moment method, SMM). The zeroth moment $\mu_0$
encodes the spectrum normalization, while higher moments encode the shape information, starting from $\mu_1$ 
(the average $\beta$ energy).
It turns out that isolines of constant $\mu_n$, as obtained by equating theoretical and experimental moments,
define conic sections in the $(r,\,s)$ plane, that for a perfect data-theory match would intersect in 
a unique point \cite{Kostensalo:2023xzu}. The crossing(s) of
the $\mu_0$ ellipse with the $\mu_1$ 
hyperbola, plus the local spread of $\mu_n$ intersections, allow to
capture  more elaborate results obtained from detailed fits to $^{113}$Cd data
\cite{Kostensalo:2020gha}.

In this work{\color{black},} we have applied the SMM  to the analysis of the $^{115}$In decay spectrum
within the same nuclear models (IBFM-2, MQPM, ISM) considered in \cite{Kostensalo:2023xzu}, by using three
independent $^{115}$In input datasets, dubbed as AC24 \cite{Pagnanini:2024qmi}, LE22 
\cite{Leder:2022beq}, and PF79 \cite{Pfeiffer:1979zz} 
{\color{black} (Fig.~\ref{Fig_01}).} These datasets, characterized by different thresholds,
show some relative differences in normalization and shape (Figs.~\ref{Fig_01} and \ref{Fig_02}) that,
not surprisingly, lead to variant results in terms of $(r,\,s)$. The analysis allows to highlight both separate and common underlying features, that should be further
investigated by future measurements and possibly by more refined nuclear models.
Our main findings are the following:
\begin{itemize}

\item  For the AC24 dataset, the SSM allows to recover, within few percent, the $(r,\,s)$ results obtained 
from detailed data fits with the enhanced spectrum-shape method 
\cite{Pagnanini:2024qmi}, showing the usefulness and simplicity of our approach (Fig.~\ref{Fig_03}). 
For the preferred $(r,\,s)$ values at $s>0$,
the sub-threshold behavior of the spectrum is expected to be either 
increasing or slightly decreasing  (Fig.~\ref{Fig_04}).

\item For the LE22 dataset, the SMM allows to reasonably recover the same $r$ values 
as from full data fits at fixed $s$ \cite{Leder:2022beq}. For free $s$, the preferred 
$r$ values are somewhat different from those of AC24, and $s<0$ is favored (Fig.~\ref{Fig_05}). 
The sub-threshold  spectrum shape for $s<0$ 
is expected to be generally decreasing (Fig.~\ref{Fig_06}). 

\item For the PF79 data analysis (Fig.~\ref{Fig_07}), we can recover only roughly the main features
of the input spectrum, namely, the peak at intermediate energy and the rapid drop at low energy, especially for $s<0$  (Fig.~\ref{Fig_08}).

\item In any model (IBFM-2, MQPM, ISM) it appears that $g_{\rm A}$ quenching ($r<1.276$) is generally needed
to match the spectrum normalization and shape for any input dataset (AC24, LE22, PF79). 
However, there is still large scatter in the admissible $(r,\,s)$ values for different data or models, with $r$ ranging between $\sim 0.8$ and $\sim 1.2$, and $s$ flipping between $\pm O(1)$. 

\item There is also ambiguity in the low-energy behavior of the spectrum, that may be increasing, stationary or decreasing in various nuclear model descriptions of different data
for variable $(r,\,s)$. Cases with $s>0$ ($s<0$) appear to be 
generally associated to an increase (decrease) of the spectra at low energy. Stabilizing the sign (and 
size) of $s$ would lead to more definite indications on $r$, since both parameters are strongly
correlated along the moment
isolines.

\item The inferred half-life $t_{1/2}$, being inversely proportional to the total spectrum area, is affected
by both normalization and shape issues. The poorly known behavior of the spectrum at low
energy affects the area and thus also $t_{1/2}$. We note that, at present,
independent experimental estimates of $t_{1/2}$ in the AC24, LE22{\color{black},} and PF79 datasets  imply somewhat
different sub-threshold extrapolations, and thus should not be statistically averaged {\color{black} in principle}.

\item A better understanding of current spectral differences and variants, together with their implications for 
allowed $(r,\,s)$ values,  low-energy spectral shapes, and half-life estimates  in various nuclear models, would 
greatly benefit from further decay data at the lowest possible experimental threshold.  
\end{itemize}

We hope that our SMM findings 
may promote further research on $^{115}$In $\beta$ decay 
and
on other $g_{\rm A}$-sensitive decays or weak processes. 
Concerning $^{115}$In decay,
we understand that a reanalysis of acquired AC24 data (or a further run with new data) in the ACCESS experiment,
pushing the threshold below 80~keV, is under consideration \cite{PrivatePagnanini}.  
Concerning other processes, of particular 
interest is the spectrum of second-forbidden $\beta$ decay of $^{99}$Tc, that has been
recently measured with high accuracy at low threshold in \cite{Paulsen:2023qnm}, and whose implications for
quenched, unquenched{\color{black},} or even enhanced values of $g_{\rm A}$ are being actively investigated 
\cite{Paulsen:2023qnm,Ramalho:2023zlt,Ramalho:2023voq,DeGregorio:2024ivh}. 
{\color{black} New measurements of forbidden decay spectra with a silicon drift detector technique are also being planned 
\cite{Nava:2024vaz}.}
From a 
theoretical viewpoint, it is hoped that the  challenges set by the experimental results 
may also promote improved (or new) approaches to nuclear models, possibly leading to
a closer match with data and to a better interpretation of the  
model parameter values.
We think that the simplicity of the spectral moment method, applied in 
\cite{Kostensalo:2023xzu} to $^{113}$Cd and herein to $^{115}$In decay, 
may provide useful insights also in {\color{black} phenomenological} analyses of $^{99}$Tc decay and other weak-interaction spectra.

\acknowledgments

The authors are grateful to Lorenzo Pagnanini and Stefano Ghislandi, and to Alexander Leder, for very useful communications about the AC24 and LE22 datasets, respectively. 
The work of E.L.\ and A.M.\  was partially supported by the research grant number 2022E2J4RK ``PANTHEON: Perspectives in Astroparticle and
Neutrino THEory with Old and New messengers,'' under the program PRIN 2022 funded by the Italian Ministero 
dell'Universit\`a e della Ricerca (MUR) and by the European Union -- Next Generation EU, as well as by the Theoretical Astroparticle Physics (TAsP) initiative of the Istituto Nazionale di Fisica Nucleare (INFN).


\end{document}